\documentclass[conference,letter]{IEEEtran}
\IEEEoverridecommandlockouts

\usepackage{cite}
\usepackage{amsmath,amssymb,amsfonts}
\usepackage{algorithmic}
\usepackage{graphicx}
\usepackage{textcomp}
\usepackage{xcolor}
\usepackage{comment}
\usepackage{mathtools}

\usepackage{float}
\usepackage[short]{optidef}
\usepackage{cleveref}
\usepackage{algorithm}
\usepackage{algorithmic}
\usepackage{subcaption}
\usepackage{graphicx}
\usepackage[acronym]{glossaries}

\newcommand{\numberofusers}{K} 
\newcommand{\user}{k}
\newcommand{\userSet}{\mathcal{\numberofusers}}
\newcommand{\datasetsize}{D}
\newcommand{\dataset}{\mathcal{\datasetsize}}
\newcommand{\scheduled}{s}
\newcommand{\scheduledvec}{\boldsymbol{\scheduled}}

\newcommand{\numberofresourceblocks}{B}
\newcommand{\resourceblock}{b}

\newcommand{\inputsvec}{\boldsymbol{x}}

\newcommand{\channel}{h}
\newcommand{\channelEst}{\hat{\channel}}
\newcommand{\resourceallocation}{\lambda}
\newcommand{\resourceallocationvec}{\boldsymbol{\lambda}}
\newcommand{\resourceallocationMat}{\boldsymbol{\Lambda}}
\newcommand{\sinr}{\gamma}
\newcommand{\sinrEst}{\hat{\sinr}}
\newcommand{\sinrTH}{\sinr_{0}}
\newcommand{\e}{\varepsilon}
\newcommand{\auxilary}{\nu}
\newcommand{\auxilaryX}{l}
\newcommand{\q}{q}
\newcommand{\qX}{g}
\newcommand{\trade}{\phi}
\newcommand{\coeffKnowledge}{\varphi}

\newcommand{\lossfunc}{f}
\newcommand{\rate}{r}
\newcommand{\power}{p}
\newcommand{\noise}{N_0}
\newcommand{\interference}{I}
\newcommand{\Dual}{\psi}

\newcommand{\re}{\varrho}
\newcommand{\dualvar}{\theta}
\DeclareMathOperator*{\argmax}{argmax}
\DeclareMathOperator*{\argmin}{argmin}
\newcommand{\model}{w}
\newcommand{\modelvec}{\boldsymbol{\model}}
\newcommand{\knowledge}{j}
\newcommand{\knowledgevec}{\boldsymbol{\knowledge}}
\newcommand{\gpr}{J}
\newcommand{\covariance}{c}
\newcommand{\covarianceVec}{\covariance}
\newcommand{\covarianceMat}{\boldsymbol{C}}
\newcommand{\gprLength}{\zeta_1}
\newcommand{\gprPeriod}{\zeta_2}

\newcommand{\trainingDuration}{T}

\newcommand{\transpose}{^\dagger}
\newcommand{\optimal}{^\star}
\newcommand{\one}{\mathbf{1}}
\newcommand{\zero}{\mathbf{0}}

\DeclareMathOperator*{\sumsum}{\sum\sum}

\newcommand{\vanilaFL}{\text{IDEAL}}
\newcommand{\propPerfect}{\text{QAW}}
\newcommand{\propImperfect}{\text{QAW-GPR}}
\newcommand{\basePaper}{\text{QUNAW}}
\newcommand{\baseRand}{\text{RAND}}

\usepackage{soul}
\usepackage{xcolor}
\def\showcomments{1}
\newcommand{\sps}[2]{%
    \if\showcomments#1%
    \textcolor{red}{[#2]}%
    \else
    \textcolor{red}{[\st{#2}]}%
    \fi}

\usepackage{balance}\newacronym{rb}{RB}{resource block}

\newacronym{fl}{FL}{federated learning}

\newacronym{csi}{CSI}{channel state information}

\newacronym{gpr}{GPR}{Gaussian process regression}

\begin{document}

\title{Federated Learning under Channel Uncertainty: Joint Client Scheduling and Resource Allocation}
\author{\IEEEauthorblockN{Madhusanka Manimel Wadu, 
	Sumudu Samarakoon, 
	and Mehdi Bennis}
\IEEEauthorblockA{\textit{Centre for Wireless	Communications (CWC), University of Oulu,
Finland} \\
\{madhusanka.manimelwadu,~sumudu.samarakoon,~mehdi.bennis\}@oulu.fi 
}
\thanks{This research was supported by the Kvantum institute strategic project SAFARI, CARMA, MISSION, NOOR, SMARTER, and the Academy of Finland 6Genesis Flagship project under grant 318927.}
}\maketitle
\begin{abstract}
In this work, we propose a novel joint client scheduling and \gls{rb} allocation policy to minimize the loss of accuracy in \gls{fl} over wireless compared to a centralized training-based solution, under imperfect \gls{csi}.
%
%
First, the problem is cast as a stochastic optimization problem over a predefined training duration and solved using the Lyapunov optimization framework.
In order to learn and track the wireless channel,  a \gls{gpr}-based channel prediction method is leveraged and incorporated into the scheduling decision.
The proposed scheduling policies are evaluated via numerical simulations, under both perfect and imperfect \gls{csi}.
Results show that the proposed method reduces the loss of accuracy up to $25.8\,\%$ compared to state-of-the-art client scheduling and RB allocation methods.
\end{abstract}

\begin{IEEEkeywords}
Federated learning, channel prediction, client scheduling, Gaussian process regression
\end{IEEEkeywords}

\glsresetall 

\section{Introduction}
\label{int}

The proliferation of a new breed of autonomous devices and mission-critical applications  sparked a huge interest in machine learning at the network edge, coined \emph{Edge ML} \cite{park2018wireless}.
In  edge ML,  training data is unevenly distributed over a large
number of devices,  and  every device has a tiny
fraction  of  the  data. 
Moreover,  devices  communicate  and
exchange their locally trained models instead of exchanging their private data. 
Among the most popular edge ML model training is \emph{federated learning}, in which the goal is to train a high quality ML model in a decentralized manner, based on local model training and client-server communication \cite{ konevcny2016federated,park2018wireless,samarakoon2018federated}.
Except a handful of works \cite{chen2018lag,nishio2019client,yang2019scheduling,chen2019joint}, the vast majority of the existing literature assumes ideal client-server communication conditions, overlooking channel dynamics and uncertainties.
In \cite{ chen2018lag}, communication overhead is reduced by using the \emph{lazily aggregate gradients (LAG)} based on reusing outdated gradient updates.
Authors in \cite{nishio2019client} propose a client-scheduling algorithm for FL considering communication and computation delays without accounting the loss of training accuracy.
In \cite{yang2019scheduling}, authors study the impact of conventional scheduling policies (e.g., random, round robin, and proportional fair) on the accuracy of FL over wireless networks without deriving the optimal scheduling policy.
In \cite{chen2019joint}, the training loss of FL is minimized by joint power allocation and client scheduling 
as a series of independent problems defined per model exchange iteration.
It can be noted that the communication aspects in FL such as optimal client scheduling and resource allocation over the entire training duration, even under the absence of the perfect channel state information (CSI) are neglected in all the aforementioned works.

The main contribution of this paper is \textbf{a novel joint client-scheduling and RB allocation policy for FL under imperfect CSI.}
We consider a set of clients  that communicates with a server over wireless links to train a neural network (NN) model within a predefined training duration.
First, we derive an analytical expression for the loss of accuracy in FL with scheduling compared to a centralized training method.
Under imperfect CSI, we adopt a Gaussian process regression (GPR)-based method to learn and track the wireless channel and quantify the information on the unexplored CSI over the network.
Then, we cast the client scheduling and RB allocation problem to minimize the loss of FL accuracy while 
acquiring as much information about the unexplored CSI
under communication constraints.
Due to the stochastic nature of the aforementioned problem, we resort to the drift-plus-penalty (DPP) technique from the Lyapunov optimization framework \cite{neely2010stochastic}.
Simulation results show that the proposed methods achieve up to 25.8\,\% reduction in loss of accuracy compared to state-of-the-art client scheduling and RB allocation methods.

The rest of this paper is organized as follows.
Section \ref{pr} presents the system model
and formulates the NN model training over wireless links, under imperfect CSI.
In Section \ref{sec:solution}, the problem is first recast in terms of loss of accuracy with scheduling compared to a centralized training method.
Then, GPR-based CSI prediction is introduced
and Lyapunov optimization is used to seek the client scheduling and RB allocation policies under both perfect and imperfect CSI.
Section \ref{res} evaluates the proposed scheduling policies. Finally, conclusions are drawn in Section \ref{conclu}.
\section{System Model and Problem Formulation}
\label{pr}
Consider a system consisting a set $ \mathcal{\numberofusers} $ of $ \numberofusers $ clients that communicate with a server over wireless.
Therein, the $ \user $-th client has a private dataset $ \mathcal{\dataset}_\user $ of size $\datasetsize_{\user} $, which is a partition of the global dataset $ \mathcal{\dataset} $ of size
$ \datasetsize = \sum_{\user} \datasetsize_{\user} $.
A set $ \mathcal{\numberofresourceblocks} $ of $ \numberofresourceblocks (\le \numberofusers $) \glspl{rb} are shared among the clients when communicating with the server.

Let $ \scheduled_{\user}(t) \in \{0,1\} $ be an indicator where $\scheduled_{\user}(t) =1$ indicates that the client $ \user $ is scheduled by the server for uplink communication at time $ t $ and $ \scheduled_{\user}(t) =0 $ otherwise. 
To schedule several clients simultaneously, one RB is allocated to each scheduled client.
Hence, we define the RB allocation vector $\boldsymbol{\resourceallocation}_{\user}(t) = [\resourceallocation_{\user,\resourceblock}(t)]_{\forall\resourceblock\in\mathcal{\numberofresourceblocks}}$ for client $\user$ with $ \resourceallocation_{\user,\resourceblock}(t) =1 $ when RB $ \resourceblock $ is allocated to client $\user$ at time $t$, and $ \resourceallocation_{\user,\resourceblock}(t) = 0 $ otherwise.
The client scheduling and RB allocation are constrained as follows:
\begin{equation}\label{eq:scheduleresource}
    \scheduled_{\user}(t) 
    \leq
    \one\transpose \boldsymbol{\resourceallocation}_{\user}(t)
    \leq 
    1
    \quad \forall \user, t.
\end{equation}
where $\one \transpose$ refers to the transpose of all one vector.
The rate at which the $k$-th client communicates with the server at time $ t $ is given by,
\begin{equation}\label{rate}
\rate_{\user}(t)  
= 
\textstyle \sum_{\resourceblock \in \mathcal{\numberofresourceblocks}}
\resourceallocation_{\user,\resourceblock}(t)
\log_2 \big( 1 + 
\frac{\power |\channel_{\user,\resourceblock}(t)|^2} { \interference_{\user,\resourceblock}(t) + \noise }
\big),
\end{equation}
where $\power$ is a fixed transmit power of client $ \user $, $ \channel_{\user,\resourceblock} (t) $ is the channel between client $ \user$ and the server over RB  $ \resourceblock$ at time $ t $, $\interference_{\user,\resourceblock}(t) $ represents the uplink interference on client $\user$ from other client over RB $\resourceblock$, and $ \noise $ is the noise power spectral density.

Under imperfect CSI, the channels need to be estimated via sampling prior to transmission.
The channel sampling data at time $t$ is collected per RB allocation over the transmissions throughout $\{1,\ldots,t-1\}$, then the future channel is inferred using the past observations as $\channelEst(t) = \gpr\big(t, \{t_n,\channel(t_n)\}_{n\in\mathcal{N}(t)} \big)$.
Here, $t_n$ is a sampling time instant and the set $\mathcal{N}(t)$ consists of sampling indices until time $t$, i.e., $n\in\mathcal{N}(t)$ is held only if $\scheduled(t_n)=1$ and $t_n<t$.  
With the estimated channels, a successful communication between a scheduled client and the server is defined by satisfying a target minimum rate.
Therefore, according to \eqref{eq:scheduleresource}, the rate constraint can be imposed per RB allocation in terms of a target signal to interference plus noise ratio (SINR) $\sinrTH$ as follows:
\begin{equation}\label{eq:allocationindicator}
\resourceallocation_{\user,\resourceblock}(t) 
\leq 
\mathbb{I} \big( \sinrEst_{\user,\resourceblock}(t) \geq \sinrTH \big)
\quad \forall \user,\resourceblock,t,
\end{equation}
where
$\sinrEst_{\user,\resourceblock}(t) = \frac{\power |\channelEst_{\user,\resourceblock}(t)|^2} { \interference_{\user,\resourceblock}(t) + \noise }$ 
and the indicator $\mathbb{I}(\sinrEst \geq \sinrTH) =1$ only if $\sinrEst \geq \sinrTH$.

The aim of model training is to minimize a regularized loss function $F(\modelvec,\mathcal{\dataset}) = \frac{1}{\datasetsize}
\sum_{\boldsymbol{\inputsvec}_i\in\mathcal{\dataset}}\lossfunc(\boldsymbol{\inputsvec}_i\transpose\modelvec ) + \xi \re( \modelvec)$ by fitting a weight vector $\modelvec$ that is known as the \emph{model} over the global dataset within a predefined communication duration $\trainingDuration$.
Here, $\lossfunc(\cdot)$, $\re(\cdot)$, and $\xi$ are the loss function, the regularization function, and the regularization coefficient, respectively. 
Due to the limitations of communication and privacy, we adopt FL as model training technique to derive the optimal weights that minimize $F(\modelvec,\mathcal{\dataset})$.
In FL, each client computes a local model over its local dataset and shares the local model with the server.
Upon receiving the local models from all clients, the server does model averaging, calculates the global model, which is broadcasted to all clients.

\begin{figure}[!t]
	\centering
	\includegraphics[width=1 \columnwidth]{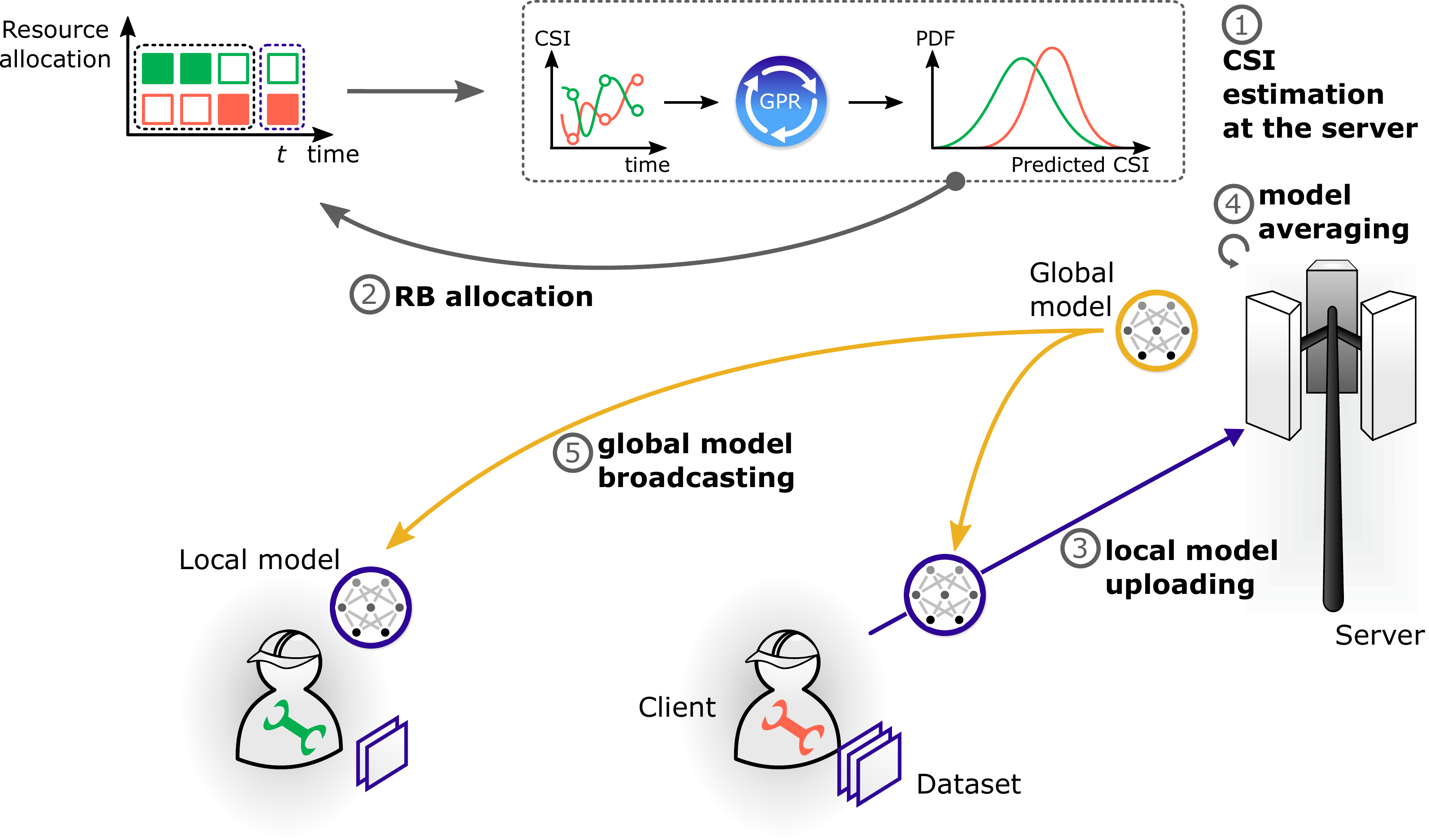}
	\caption{FL with client scheduling under limited wireless resources and imperfect CSI.}
	\label{fig:system_model}
\end{figure}

Under imperfect CSI, channels between clients and the server over each RB are predicted using their past observations prior to the communication.
With $\resourceallocation_{\user,\resourceblock}(t)=1$, the channel $\channel_{\user,\resourceblock}(t)$ is sampled and used as an observation in the future.
It means that the RB allocation and channel sampling are carried out simultaneously.
In this regard, we define the information on the channel between client $ \user $ and server at time $t$ that can be obtained by RB allocation as $\knowledgevec_\user(t) = [\knowledge_{\user,\resourceblock}(t)]_{\user\in\userSet}$.
For accurate CSI predictions, it is essential to acquire as much information about the CSI over the network \cite{karaca2012smart}.
In this view, we maximize $\sum_{\user} \knowledgevec_\user\transpose(t) \resourceallocationvec_\user(t)$ at each $t$ while minimizing the loss $F(\modelvec,\mathcal{\dataset})$.
This iterative process is carried out over a training duration of $\trainingDuration$ as illustrated in Fig. \ref{fig:system_model}.
The empirical loss minimization problem for all $t\in\{1,\ldots,\trainingDuration\}$ is formally defined as follows:
\begin{subequations}\label{eq:master_problem}
\begin{align}
\label{obj:master_problem}
\underset{\modelvec(t), \scheduledvec(t), \resourceallocationMat(t), \forall t}{\text{minimize}} 
& F\big( \modelvec(T), \mathcal{\datasetsize} \big) 
- 
\textstyle 
\frac{\coeffKnowledge}{\trainingDuration}
\sum_{\user,t} \knowledgevec_\user\transpose(t) \resourceallocationvec_\user(t)
\\
\label{cns:model_constraints}
\text{subject to\hphantom{00}} 
& \eqref{eq:scheduleresource}\text{-}\eqref{eq:allocationindicator}, \\
\label{cns:OFDMA}
& \boldsymbol{A}\resourceallocationMat\transpose(t) \preceq \one,\\
\label{cns:RB_availability}
& \one\transpose\scheduledvec(t) \leq \numberofresourceblocks,\\
\label{cns:boolean}
& \scheduledvec(t) \in \{0,1\}^\numberofusers,\resourceallocationvec_\user(t) \in \{0,1\}^\resourceblock,\\ 
\label{cns:local_SGD}
& \modelvec_{\user}(t) = {\small \argmin_{\modelvec'} F( \modelvec'| \modelvec(t-1),\dataset_\user}),\\ 
\label{cns:model_update}
& \textstyle \modelvec(t) = \sum_\user \frac{\datasetsize_\user}{\datasetsize} \scheduled_\user(t) \modelvec_\user(t),  
\end{align}
\end{subequations}
where 
$\resourceallocationMat\transpose(t) = [\resourceallocationvec_\user\transpose(t)]_{\user\in\userSet}$,
$\coeffKnowledge(>0)$ controls the impact of the information exploration,
and
$\boldsymbol{A}$ is a $\numberofresourceblocks\times\numberofusers$ all-one matrix.
The orthogonal channel allocation in \eqref{cns:OFDMA} ensures collision-free client uplink transmission with $I_{\user,\resourceblock}(t)=0$ and constraint \eqref{cns:RB_availability} defines the maximum allowable clients to be scheduled due to the limitation in the RB availability.
The stochastic gradient decent (SGD) based local model calculation at client $\user$ is defined in \eqref{cns:local_SGD}.

\section{Optimal client-Scheduling and RB Allocation Policy via Lyapunov Optimization}\label{sec:solution}

It can be noted that the optimization problem \eqref{eq:master_problem} is coupled over all clients.
Hence, in what follows, the discussion of decoupling \eqref{eq:master_problem} over clients and the server, and then deriving the optimal client scheduling and RB allocation policy.

\subsection{Decoupling \eqref{eq:master_problem} via dual formulation}
Let us consider an ideal unconstrained scenario where the server gathers the entire data samples and trains the global model in a \emph{centralized} manner.
Let $F_0 = \min_{\modelvec} F(\modelvec,\dataset)$ be the minimum loss under centralized training.
By the end of training duration $\trainingDuration$, we define the gap between the studied FL under communication constraints and centralized training as $\e(\trainingDuration) = F\big( \modelvec(T), \mathcal{\datasetsize} \big) - F_0$.
Here, $\e(\trainingDuration)$ is the loss of FL with scheduling compared to centralized training.
Note that minimizing \eqref{obj:master_problem} remains unchanged by minimizing the gap $\e(\trainingDuration)$ under the same set of constraints.

To analyse the loss of FL with scheduling, we consider the dual function of \eqref{obj:master_problem} with the dual variable $\boldsymbol{\dualvar}=[\dualvar_1,\ldots,\dualvar_{\datasetsize}]$ and $\boldsymbol{X}=[\boldsymbol{X}_\user]_{\user\in\userSet}$ with $\boldsymbol{X}_\user = [ \inputsvec_i ]_{i=1}^{\datasetsize_\user}$ as follows:
	\begin{align}
		\nonumber
		\Dual(\boldsymbol{\dualvar})
		&=
		\min_{\boldsymbol{w}, \boldsymbol{z}} 
		\Big(  
		\textstyle \sum_{\boldsymbol{\inputsvec}_i\in\mathcal{\datasetsize}}
		\frac{1}{\datasetsize}\lossfunc_i( \boldsymbol{\inputsvec}_i^T\boldsymbol{w} ) 
		+ \xi \re ( \boldsymbol{w})
		+ \frac{ \boldsymbol{\dualvar}^T(\boldsymbol{\dualvar} - \boldsymbol{z}) }{\datasetsize}  \Big) \\	
		&= 
		\textstyle 
		- \sum_{k=1}^{\numberofusers}  \sum_{i=1}^{\datasetsize_\user} \frac{1}{\datasetsize} \lossfunc_i^* (-\dualvar_i)
		-\xi \re^*(\boldsymbol{v}),
		\label{eq:conjugates}
	\end{align}
\normalsize
where 
$\boldsymbol{v} = {\boldsymbol{X}\boldsymbol{\dualvar}}/{\xi \datasetsize}$,
$ \boldsymbol{z}=\boldsymbol{X}^T\boldsymbol{w}$ is a newly introduced variable,
and
$ \lossfunc^*(.) $, $ \re^*(.) $ are the conjugate functions of $ \lossfunc(.) $ and $ \re(.) $, respectively.
With the dual formulation the relation between the primal and dual variables is $\boldsymbol{w} = \nabla \re ^*(\boldsymbol{v})$ \cite{yang2019scheduling}.
Based on the dual formulation, the loss of FL with scheduling is $\e(\trainingDuration) = \Dual_0 - \Dual\big( \boldsymbol{\dualvar}(T) \big)$ where $\Dual_0$ is the maximum dual function value obtained from the centralized method.

Note that the first term of \eqref{eq:conjugates} decouples per client and thus, can be computed locally.
In contrast, the second term in \eqref{eq:conjugates} cannot be decoupled per client.
To compute $\re^*(\boldsymbol{v})$, first, each client $ \user $ locally computes $ \Delta \boldsymbol{v}_\user(t)  = \frac{1}{\xi \datasetsize} \boldsymbol{X}_{\user} \Delta \boldsymbol{\dualvar}_{\user}(t) $ at time $ t $.
Here, $ \Delta \boldsymbol{\dualvar}_k(t) $ is the change in dual variable $ \boldsymbol{\dualvar}_k(t) $ for client $ k $ in the time $ t $ given as below,
\begin{multline}\label{eq:delta_theta}
	\Delta \boldsymbol{\dualvar}_{\user}(t) 
	\approx
	\textstyle \argmax_{ \boldsymbol{\delta} \in \mathbb{R}^{\datasetsize_\user}} 
	\Big( 
	-\frac{1}{\datasetsize} \one\transpose  [ \lossfunc_i^* (-\boldsymbol{\dualvar}_{\user}(t) - \boldsymbol{\delta}) ]_{i =1}^{\datasetsize_\user} \\
	\textstyle
	- \frac{\xi}{\numberofusers} \re^*\big(\boldsymbol{v}(t) \big) 
	-  \frac{1}{\datasetsize} \boldsymbol{\delta}\transpose \boldsymbol{X}_{k} \re^*\big(\boldsymbol{v}(t)\big) 
	- \frac{\eta/\xi}{2 \datasetsize^2} \| \boldsymbol{X}_{\user} \boldsymbol{\delta}\|^2
	\Big),
\end{multline}
where $\eta$ depends on the partitioning of the $\dataset$ \cite{hiriart2004fundamentals}.
It is worth noting that $\Delta \boldsymbol{\dualvar}_k(t)$ in \eqref{eq:delta_theta} is computed based on the previous global value $\boldsymbol{v}(t)$ received by the server.
Then, the scheduled clients upload $(\Delta \boldsymbol{v}_\user(t),\Delta \boldsymbol{\dualvar}_{\user}(t))$ to the server.
Following the dual formulation, the model aggregation and update in \eqref{cns:model_update} at the server is modified as follows:
\begin{subequations}\label{eqn:model_update_new}
\begin{align}
\boldsymbol{v}(t+1)
&\coloneqq   \textstyle 
\boldsymbol{v}(t) + \sum_{\user \in \mathcal{\numberofusers}} \scheduled_{\user}(t) \Delta \boldsymbol{v}_\user(t), \label{eq5-1} \\
\boldsymbol{\dualvar}(t+1) 
&\coloneqq \textstyle 
\boldsymbol{\dualvar}(t) + \sum_{\user \in \mathcal{\numberofusers}} \frac{1}{\numberofusers} \scheduled_{\user}(t) \Delta \boldsymbol{\dualvar}_{\user}(t). \label{eq5-2}
\end{align}
\end{subequations}
Using \eqref{eq5-1}, the server computes the coupled term $\re^*\big(\boldsymbol{v}(t+1)\big)$ in \eqref{eq:conjugates}.

Note that from the $t$-th update, $\Delta \boldsymbol{\dualvar}_k(t)$ in \eqref{eq:delta_theta} maximizes $\Delta\Dual\big(\boldsymbol{\dualvar}_\user(t)\big)$, which is the change in dual function $\Dual\big(\boldsymbol{\dualvar}(t)\big)$ corresponding to the  of client $\user$. 
Let $\boldsymbol{\dualvar}\optimal_\user(t)$ be the local optimal dual variable at time $t$, in which $\Delta\Dual\big(\boldsymbol{\dualvar}\optimal_\user(t)\big) \geq \Delta\Dual\big(\boldsymbol{\dualvar}_\user(t)\big)$ is held.
Then for a given accuracy  $\beta_{\user}(t) \in (0,1) $ of local SGD updates, the following condition is satisfied:
\begin{equation}\label{beta}
	\textstyle \frac{  \Delta \Dual_\user\big(\Delta \boldsymbol{\dualvar}\optimal_{\user} (t)\big)  - \Delta \Dual_\user\big(\Delta \boldsymbol{\dualvar}_{\user} (t)\big)  }{ \Delta \Dual_\user\big(\Delta \boldsymbol{\dualvar}_{\user} (t)\big)  - \Delta \Dual_\user(0) } \leq \beta_{\user}(t),
\end{equation}
where
$ \Delta \Dual_\user(0) $ is the change in $ \Dual $ with a null update from $ k $-th client.
For simplicity, we assume that $\beta_{\user,t} = \beta$ for all $\user\in\userSet$ and $t$, hereinafter.
With \eqref{beta}, the gap between FL with scheduling and the centralized method is bounded as follows \cite[Appendix B]{yang2019scheduling}:
\begin{equation}\label{eq:gap}
	\e(\trainingDuration) \leq \datasetsize \Big( 1-(1-\beta) 
	\textstyle \sumsum_{t\leq\trainingDuration}^{\user\leq\numberofusers} \frac{\datasetsize_\user}{\trainingDuration\datasetsize} \scheduled_\user(t)
	\Big)^\trainingDuration.
\end{equation}
This yields that the minimization of $\e(\trainingDuration)$ can be achieved by minimizing its upper bound defined in \eqref{eq:gap}.
Henceforth, the equivalent form of \eqref{eq:master_problem} is given as follows:
\begin{subequations}\label{eq:modified_problem}
\begin{align}
\nonumber
\underset{[\Delta \boldsymbol{\dualvar}_{\user}(t)]_{\user}, \scheduledvec(t), \resourceallocationMat(t), \forall t}{\text{minimize}} 
& \datasetsize \Big( 1- (1-\beta)
	\textstyle \sum_{t,\user} \frac{\datasetsize_\user}{\trainingDuration\datasetsize} \scheduled_\user(t)
	\Big)^\trainingDuration \\
	\label{obj:modified_problem}
	& \textstyle \qquad
	- \frac{\coeffKnowledge}{\trainingDuration}
	\sum_{\user,t} \knowledgevec_\user\transpose(t) \resourceallocationvec_\user(t)  \\
\label{cns:all_modified}
\text{subject to \hphantom{000}} 
& \eqref{cns:model_constraints}\text{-}\eqref{cns:boolean}, \eqref{eq:delta_theta}, \eqref{eqn:model_update_new}.
\end{align}
\end{subequations}

\subsection{GPR-based metric for information on unexplored CSI}\label{subsec:gpr}

For CSI predictions, we use GPR with a Gaussian kernel function to estimate the nonlinear relation of $\gpr(\cdot)$ by assuming that it follows a Gaussian process (GP) as a prior.
In this view, for a finite data set $\{t_n,\channel(t_n)\}_{n\in\mathcal{N}}$, the aforementioned GP becomes a multi-dimensional Gaussian distribution, with a zero mean and covariance $\covarianceMat = [\covariance(t_m,t_n)]_{m,n\in\mathcal{N}}$ given by,
\begin{equation}\label{eq:gpr_covar}
    \covariance(t_m,t_n) = 
    \textstyle \exp \Big( 
    -\frac{1}{\gprLength} \sin^2 \big(\frac{\pi}{\gprPeriod} (t_m-t_n) \big)
    \Big),
\end{equation}
where $\gprLength$ and $\gprPeriod$ are the length and period hyper-parameters, respectively \cite{xing2015gpr}.
Henceforth, the CSI prediction at time $t$ and its uncertainty/variance is given by \cite{PrezCruz2013GaussianPF},
\begin{eqnarray}
    \label{eq:gpr_mean}
    &\channelEst(t) 
    = \covarianceVec\transpose(t)\covarianceMat^{-1}[\channel(t_n)]_{n\in\mathcal{N}}, \\
    \label{eq:gpr_var}
    &\knowledge(t) 
    = \covariance(t,t) - \covarianceVec\transpose(t)\covarianceMat^{-1}\covarianceVec(t),
\end{eqnarray}
where
$\covarianceVec(t) = [\covariance(t,t_n)]_{n\in\mathcal{N}}$.
Note that the client and RB dependence is omitted in the discussion above for notation simplicity.
Here, the channel estimation is given in \eqref{eq:gpr_mean}.
The CSI uncertainty is used as the information $\knowledge(t)$, in which exploring highly uncertain channels provides more insight.
It is worth noting that under \emph{perfect CSI} $\channelEst(t) = \channel(t)$ and $\knowledge(t)=0$.

\subsection{Joint client scheduling and RB allocation}\label{subsec:policy}

Due to the time average objective in \eqref{obj:modified_problem}, the problem \eqref{eq:modified_problem} becomes a stochastic optimization problem defined over $t = \{1,\ldots,\trainingDuration\}$.
Therefore, we resort to the \emph{drift plus penalty} (DPP) technique in Lyapunov optimization framework to derive the optimal scheduling policy \cite{neely2010stochastic}.
Therein, Lyapunov framework allows us to transform the original stochastic optimization problem into a series of optimizations problems that are solved at each time $t$, as discussed next.

First, we denote $u(t) = (1-\beta)\sum_\user \scheduled_\user(t)\datasetsize_\user / \datasetsize$ and define its time average $\bar{u} = \sum_{t\leq\trainingDuration} u(t) / \trainingDuration$.
Then, we introduce auxiliary variables $ \auxilary(t)$ and $\auxilaryX(t)$ with time average lower bounds $\bar{\auxilary} \leq \bar{u} $ and $\bar{\auxilaryX} \leq \frac{1}{\trainingDuration}
	\sum_{\user,t} \knowledgevec_\user\transpose(t) \resourceallocationvec_\user(t) \leq \auxilaryX_0$, respectively.
To track the time average lower bounds, next we introduce virtual queues $ \q(t) $ and $\qX(t)$ with the following dynamics \cite{neely2010stochastic}:
\vspace{-0.4 cm}
\begin{subequations}\label{eq:queue}
\begin{eqnarray}
    &\q(t+1) = \max \big( 0 , \q(t) + \auxilary(t) - u(t) \big), \\
    &\qX(t+1) = \max \big( 0 , \qX(t) + \auxilaryX(t) -
	\sum_{\user} \knowledgevec_\user\transpose(t) \resourceallocationvec_\user(t) \big).
\end{eqnarray}
\end{subequations}
In this view, $ \eqref{eq:modified_problem}$ can be recast as follows:
\begin{subequations}\label{eq:lypunov_problem}
\begin{align}
\label{obj:lyapunov_problem}
\underset{\substack{[\Delta \boldsymbol{\dualvar}_{\user}(t)]_{\user}, \scheduledvec(t), \resourceallocationMat(t), \\ \auxilary(t), \auxilaryX(t) \forall t}}{\text{minimize}} 
& \datasetsize (1-\bar{\auxilary})^\trainingDuration 
	- \coeffKnowledge\bar{\auxilaryX} \\
\label{cns:lyapunov_constrains}
\text{subject to \hphantom{00}} 
& \eqref{cns:all_modified}, \eqref{eq:queue}, \\
\label{cns:aux_var}
& 0 \leq \auxilary(t) \leq 1-\beta \quad \forall t, \\
\label{cns:auxX_var}
& 0 \leq \auxilaryX(t) \leq \auxilaryX_0 \quad \forall t, \\
\label{cns:utilization}
& \textstyle u(t) = \sum_\user \frac{(1-\beta)\datasetsize_\user}{\datasetsize}\scheduled_\user(t) \quad \forall t.
\end{align}
\end{subequations}

The quadratic Lyapunov function of $ \q(t) $ is $ L (t) = {\big(\q(t)^2 + \qX(t)^2\big)}/{2} $.
Given $\big(\q(t),\qX(t)\big)$, the expected conditional Lyapunov one slot drift at time $ t $ is $ \Delta L = \mathbb{E}[L(t+1)-L(t)|\q(t),\qX(t)] $.
Weighted by a tradeoff parameter $ \trade(\geq 0) $, we add a penalty term,
\begin{multline}
\textstyle
\trade \big( 
\frac{\partial }{\partial \auxilary} [(1-{\auxilary})^T\datasetsize]_{\auxilary=\tilde{\auxilary}(t)} \mathbb{E}[\auxilary (t)  | \q(t)]  
- \coeffKnowledge \mathbb{E}[\auxilaryX (t)  | \qX(t)] 
\big) = \\
-\trade \Big( 
\datasetsize\trainingDuration\big(1-\tilde{\auxilary}(t)\big)^{\trainingDuration-1} \mathbb{E}[\auxilary (t)  | \q(t)]
+ \coeffKnowledge \mathbb{E}[\auxilaryX (t)  | \qX(t)] 
\Big),
\end{multline}
to obtain the Lyapunov DPP.
Here, $\tilde{\auxilary}(t) = \frac{1}{t}\sum_{\tau=1}^t \auxilary(\tau)$ and $\tilde{\auxilaryX}(t) = \frac{1}{t}\sum_{\tau=1}^t \auxilaryX(\tau)$ are the running time average of the auxiliary variables at time $t$.
Using the inequality $ \max ( 0 , x)^2 \leq x^2 $, the upper bound of the Lyapunov DPP is given by,
\begin{multline}\label{eq:DPP}
\Delta L 
-\trade \Big( 
\datasetsize\trainingDuration\big(1-\tilde{\auxilary}(t)\big)^{\trainingDuration-1} \mathbb{E}[\auxilary (t)  | \q(t)]
+ \coeffKnowledge \mathbb{E}[\auxilaryX (t)  | \qX(t)] 
\Big)
\leq \\
\textstyle  
\mathbb{E}[\q(t)\big( \auxilary(t) - u(t) \big) 
+ \qX(t)\big( \auxilaryX(t) - \sum_{\user} \knowledgevec_\user\transpose(t) \resourceallocationvec_\user(t) \big)
+ L_0 \\
-\trade \Big( 
\datasetsize\trainingDuration\big(1-\tilde{\auxilary}(t)\big)^{\trainingDuration-1} \auxilary (t) 
+ \coeffKnowledge \auxilaryX(t) \Big)
| \q(t),\qX(t)],
\end{multline}
where $L_0$ is a uniform bound on $ \big( \auxilary(t) - u(t) \big)^2/2 + \big( \auxilaryX(t) - \sum_{\user} \knowledgevec_\user\transpose(t) \resourceallocationvec_\user(t) \big)^2/2$ for all $t$.
The motivation behind deriving the Lyapunov DPP is that minimizing the upper bound of the expected conditional Lyapunov DPP at each iteration $t$ with a predefined $\trade$ yields the tradeoff between the virtual queue stability and the optimality of the solution for \eqref{eq:lypunov_problem} \cite{neely2010stochastic}.
In this regard, the stochastic optimization problem of \eqref{eq:lypunov_problem} is solved via minimizing the upper bound in \eqref{eq:DPP} at each time $t$ as follows:
\begin{subequations}\label{eq:problem_per_t}
\begin{align}
\nonumber
\underset{\scheduledvec(t), \resourceallocationMat(t), \auxilary(t), \auxilaryX(t) }{\text{maximize}} \hphantom{0}
& \textstyle
\sum_\user \big(
\frac{\q(t)(1-\beta)\datasetsize_\user }{\datasetsize}\scheduled_\user(t)
+ \qX(t)\knowledgevec_\user\transpose(t) \resourceallocationvec_\user(t)
\big) 
 \\
\label{obj:problem_per_t}
& \textstyle 
\qquad -  \alpha(t)\auxilary(t) 
- \big( \qX(t) - \trade\coeffKnowledge \big) \auxilaryX(t)
\\
\label{cns:per_t}
\text{subject to} \hphantom{00}
& \eqref{cns:model_constraints}\text{-}\eqref{cns:RB_availability}, \eqref{cns:aux_var}, \eqref{cns:auxX_var}, \\
\label{cns:relaxed_constrains}
& \zero \preceq \scheduledvec(t),\resourceallocationvec_\user(t) \preceq \one, 
\end{align}
\end{subequations}
where 
$\alpha(t) = \q(t) - \trade\datasetsize\trainingDuration\big(1-\tilde{\auxilary}(t)\big)^{\trainingDuration-1}$.
Note that the constant $\kappa$ is removed, the variable $\Delta \boldsymbol{\dualvar}_\user(t)$ with constraints \eqref{eq:delta_theta} and \eqref{eqn:model_update_new} are decoupled from \eqref{eq:problem_per_t}, and the Boolean variables are relaxed as linear variables.
Here, the objective and the constraints in \eqref{eq:problem_per_t} are affine, and the problem is a linear program (LP).
Due to the independence, the optimal auxiliary variables are derived by decoupling the \eqref{obj:problem_per_t}, \eqref{cns:aux_var}, and \eqref{cns:auxX_var} as follows:
\begin{equation}\label{eq:auxiliary_optimal}
\auxilary\optimal(t) =
\begin{cases}
1-\beta & \text{if}~ \alpha(t) \geq 0, \\
0 & \text{otherwise},
\end{cases}
\quad 
\auxilaryX\optimal(t) =
\begin{cases}
\auxilaryX_0  & \text{if}~ \qX(t) \geq \trade\coeffKnowledge, \\
0 & \text{otherwise}.
\end{cases}
\end{equation}
The optimal scheduling $\scheduledvec\optimal(t)$ and RB allocation variables $\resourceallocationMat\optimal(t)$ are found using an interior point method (IPM).
It is due to the nature of LP, the optimal solution of the relaxed problem lies on a vertex of the feasible convex hull yielding the optimal solution for the problem with the Boolean variables.
The joint client scheduling and RB allocation is summarized in Algorithm \ref{alg:schedluing_RB_allocation}.

 \begin{algorithm}[!t]
	\caption{Joint Client Scheduling and RB Allocation}
	\label{alg:schedluing_RB_allocation}
	\begin{algorithmic}[1]
		\renewcommand{\algorithmicrequire}{\textbf{Input:}}
		\renewcommand{\algorithmicensure}{\textbf{Output:}}
		\REQUIRE $ \dataset,\sinrTH, \beta, \power, \numberofresourceblocks, \xi $
		\ENSURE  $ \scheduledvec\optimal(t), \resourceallocationMat\optimal(t) $ for all $t$
		\STATE $ \q(0) =\qX(0)=0 $, $ \auxilary(0)=\auxilaryX(0) = 0 $,  $\boldsymbol{v}(0)=\zero$
		\FOR {$t = 1$ to $\trainingDuration$}
		\STATE Each client computes $\Delta\boldsymbol{\dualvar}_\user(t)$ using \eqref{eq:delta_theta}
		\STATE Channel prediction with \eqref{eq:gpr_mean}
		\STATE Calculate $\auxilary\optimal(t)$ and $\auxilaryX\optimal(t)$ using \eqref{eq:auxiliary_optimal}
		\STATE Derive $\scheduledvec\optimal(t)$ and $\resourceallocationMat\optimal(t)$ by solving \eqref{eq:problem_per_t} using an IPM
		\STATE Local model $( \Delta \boldsymbol{v}_\user(t),  \Delta \boldsymbol{\dualvar}_\user(t))$ uploading to the server
		\STATE Update $\tilde{\auxilary}(t)$, $\q(t)$ via \eqref{eq:queue}, $\boldsymbol{v}(t)$ and $\boldsymbol{\dualvar}(t)$ with \eqref{eqn:model_update_new}
		\STATE Global model $\boldsymbol{v}(t)$ broadcasting
		\STATE $t \to t+1 $
		\ENDFOR
	\end{algorithmic}
\end{algorithm}

\section{Simulation Results}
\label{res}

\begin{table}[!t]
    \caption{Simulation parameters }
    \label{tableofparameters}
    \begin{center}
        \begin{tabular}{c c||c c||c c}
        \hline
          \multicolumn{1}{c}{Parameter}
        & \multicolumn{1}{c||}{Value}
        & \multicolumn{1}{|c}{Parameter}
        & \multicolumn{1}{c||}{Value}
        & \multicolumn{1}{|c}{Parameter}
        & \multicolumn{1}{c}{Value}
        \\
        \hline
          $\sinrTH$      &  $1.2$    &$\numberofresourceblocks$   &  $6$    &$\beta$        &  $0.7$    \\
          $\gprLength$            &  2 &$\trade$       &  $1$      &$ \xi  $                      &  1 \\ 
          $\gprPeriod$  &  $5$       &$\eta  $                     &  $0.2$    &$|\mathcal{N}|$    &  $20$     \\
          $\trainingDuration$ & 100 & $\coeffKnowledge$ & 1 & $\power$ & 1 \\ \hline
        \end{tabular}
    \end{center}
\end{table}

In this section, we evaluate the proposed client scheduling method using MNIST dataset assuming $\lossfunc(\cdot)$ and $\re(\cdot)$ as cross entropy loss functions with a Tikhonov regularizer. 
For the training loss function, $\e_{\text{method}} = \text{accuracy}_{\text{centralized}} - \text{accuracy}_{\text{method}}$ is used.
Here, the \emph{centralized training} refers to the training takes place at the server with the access to the entire dataset.  
A dataset of 6000 samples consisting of equal sizes of ten classes for 0-9 digits are used over $\numberofusers=10$ clients.
To partition the training dataset over clients, we use the Zipf distribution to determine the local training dataset size. 
In other words client $\user$ owns a dataset with $\datasetsize_\user = \datasetsize\user^{-\sigma}/\sum_{\kappa\in\userSet} \kappa^{-\sigma}$.
Here, the Zipf's parameter $\sigma=0$ yields uniform data distribution over clients (600 samples per client), and increasing $\sigma$ results in heterogeneous sample sizes among clients.
In addition, the uplink transmission power is set to $\power=1\,$W and the channel follows a correlated Rayleigh distribution with mean to noise ratio equal to $\sinrTH$.
For perfect CSI, it is assumed that a single RB is dedicated for channel measurements.
The remaining parameters are presented in Table \ref{tableofparameters}.

\begin{figure}[!t]
    \begin{subfigure}{\columnwidth}
        \centering\includegraphics[width=\textwidth]{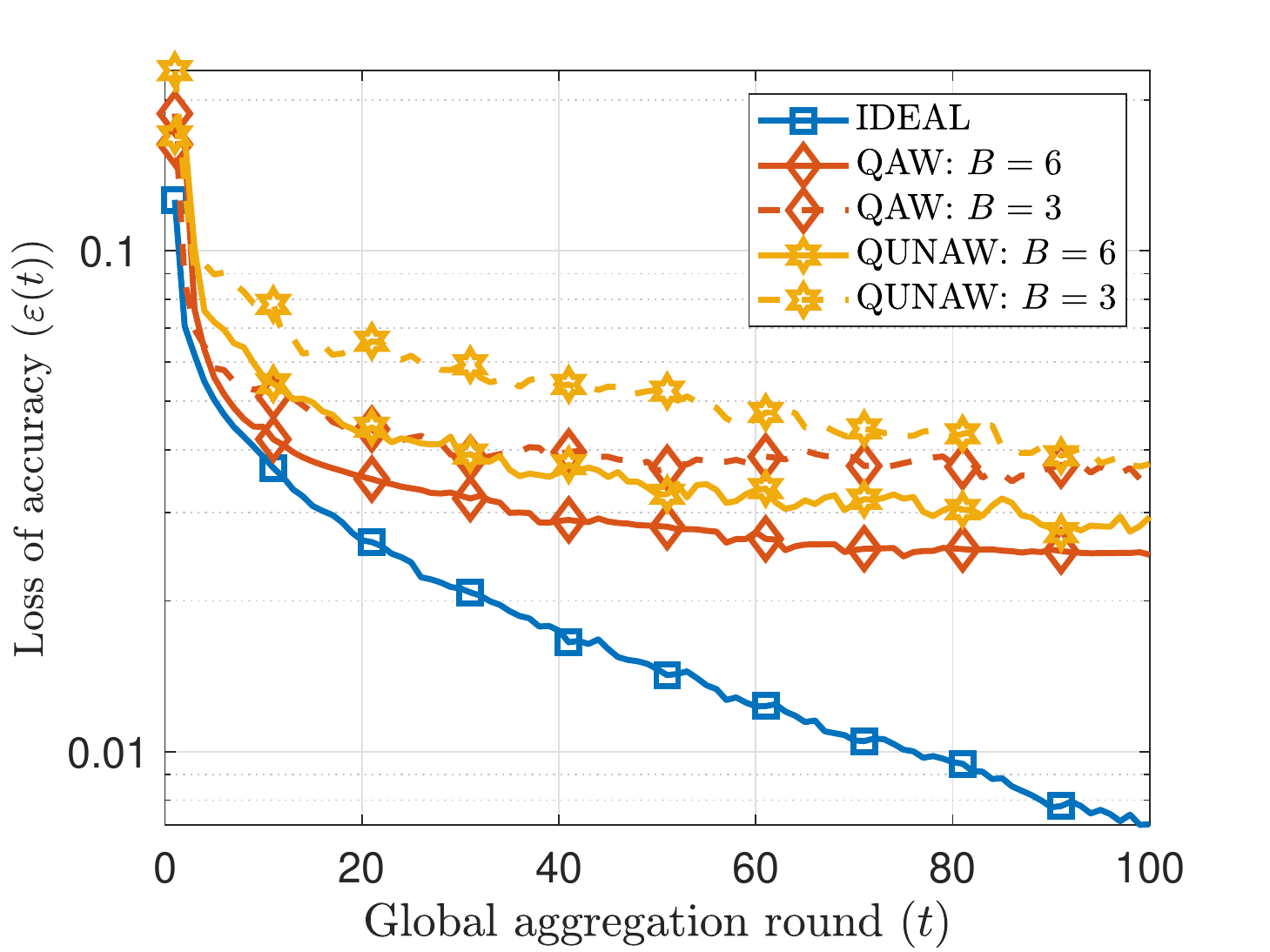}
        \caption{FL with perfect CSI and $\numberofresourceblocks=\{3,6\}$.}
        \label{fig:epsilon_perfect}
    \end{subfigure}
    \begin{subfigure}{\columnwidth}
        \centering\includegraphics[width=\textwidth]{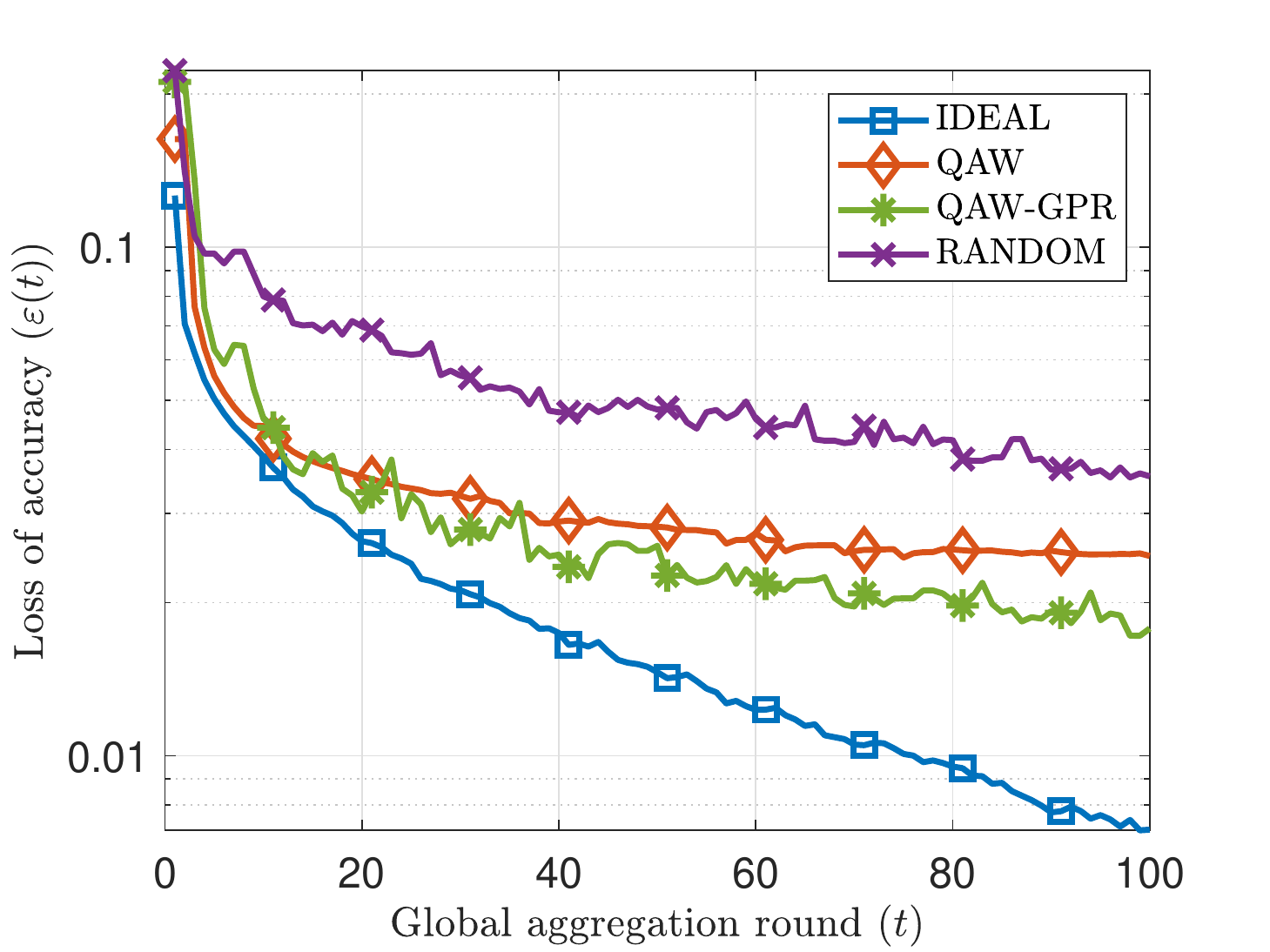}
        \caption{FL with imperfect CSI and $\numberofresourceblocks=6$.}
        \label{fig:epsilon_imperfect}
    \end{subfigure}
    \caption{Comparison of the loss of accuracy in all FL methods for each model aggregation round vs. centralized training, Zipf parameter $\sigma=1.017$. }
    \label{fig:epsillon}
\end{figure}

Under perfect CSI, we denote the proposed scheduling method as data quantity aware scheduling policy ``\propPerfect".
By deriving the optimal client scheduling and RB allocation based on the findings in \cite{yang2019scheduling}, we obtain the quantity unaware baseline ``\basePaper".
Under imperfect CSI, the proposed GPR-based channel prediction and client scheduling method is coined as ``\propImperfect".
Whereas the random scheduling baseline is denoted by ``\baseRand".
Finally, to highlight the upper bound performance, we use the \emph{vanilla FL} method \cite{konevcny2016federated} without RB constraints, which is denoted as `\vanilaFL" hereinafter.

\textbf{Loss of accuracy comparison:} Fig. \ref{fig:epsillon} compares the loss of accuracy in all FL methods at each model aggregation round with respect to centralized model training.
It can be noted that \vanilaFL{} has the lowest loss of $\e(100)=0.7$ due to the absence of communication constraints.
Under perfect CSI, Fig. \ref{fig:epsilon_perfect} plots \propPerfect{} and \basePaper{} for two different RB scenarios with $\numberofresourceblocks = 3$ and $6$.
For $\numberofresourceblocks=3$, the losses in both \propPerfect{} and \basePaper{} are almost identical.
For $\numberofresourceblocks =6$, Fig. \ref{fig:epsilon_perfect} shows that the quantity aware scheduling (\propPerfect{}) reduces the loss by $ 15.9\,\%$ compared to \basePaper{}.
Under imperfect CSI, \propImperfect{} and \baseRand{} are compared in Fig. \ref{fig:epsilon_imperfect} alongside $\vanilaFL{}$ and $\propPerfect{}$.
While \baseRand{} shows a poor performance, \propImperfect{} outperforms \propPerfect{} by reducing the loss by $28\,\%$.
The main reason for this improvement is due to the RB  utilization of all $\numberofresourceblocks=6$ RBs for scheduling clients  compared to \propPerfect{}, which allocates one RB for CSI measurements.

\begin{figure}[!t]
    \centering
    \includegraphics[width=\columnwidth]{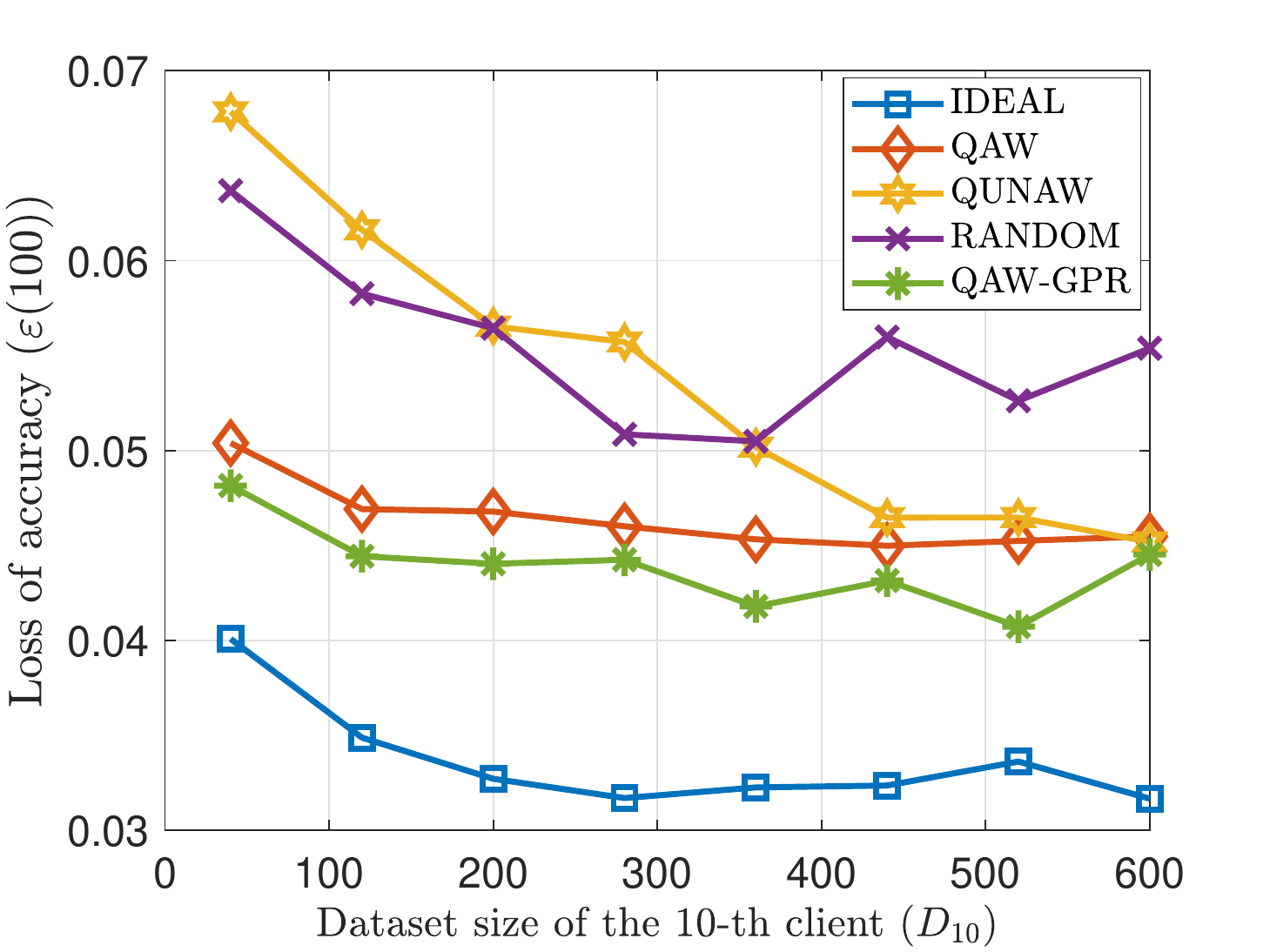}
    \caption{Comparison of the loss of accuracy per client dataset size.}
    \label{fig:accuracy}
\end{figure}

\textbf{Impact of per-client dataset size:} Fig. \ref{fig:accuracy} plots the impact of heterogeneity in the training sample size per client on the loss of accuracy.
Here, the $x$-axis represents the number of data samples per client with the minimum number of training data, i.e., the dataset size of the 10th client $\datasetsize_{10}$ as per the Zipf's distribution.
All methods exhibit higher losses in accuracy when the training samples are asymmetrically distributed over clients, i.e., for the lower $\datasetsize_{10}$.
As $\datasetsize_{10}$ is increased, the losses in accuracy are reduced.
It is also worth noting that the losses of accuracy in the proposed methods \propPerfect{} and \propImperfect{} remain almost constant for $\datasetsize_{10}>100$.
The loss reductions in \propImperfect{} over \propPerfect{} are due to the additional RB with the absence of CSI measurement.
In contrast, \basePaper{} yields higher losses when training data is unevenly distributed among clients.
The reductions of the loss in \propPerfect{} at $\datasetsize_{10}=40$ are $25.72\,\%$ and $20.87\,\%$ compared to \basePaper{} and \baseRand{}, respectively.
The reason behind the these lower losses of the proposed methods over the baselines is that client scheduling takes into account the training dataset size.
For $\datasetsize_{10}=600$, due to the equal dataset sizes per client, the accuracy loss with \propPerfect{} and \basePaper{} identical.
Therein, both \propPerfect{} and \basePaper{} exhibit about $18.4\,\%$ reduction in accuracy loss compared to \baseRand{}.

\begin{figure}[!t]
    \centering
    \includegraphics[width=\columnwidth]{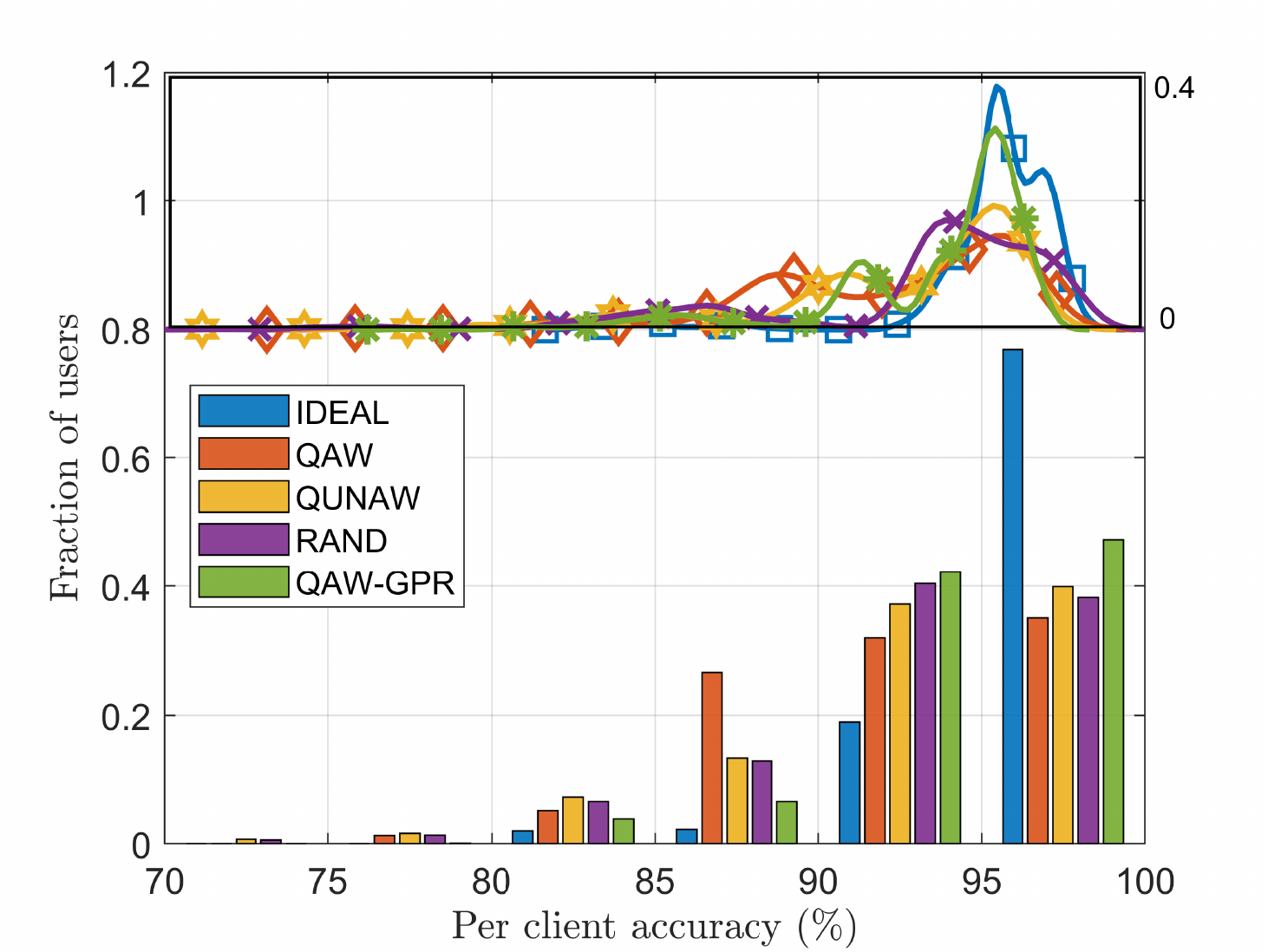}
    \caption{Fairness comparison of the training accuracy among clients, Zipf parameter $\sigma=1.071$.}
    \label{fig:peruseraccuracy}
\end{figure}

\textbf{Impact of Fairness:} Finally, in terms of fairness, the accuracy per client for different FL methods are investigated in Fig. \ref{fig:peruseraccuracy}.
Here, \vanilaFL{} exhibits the highest average training accuracy of 95.3\,\% as well as the lowest variance of 5.6 over the clients compared to all other methods.
This demonstrates that the most fairness in terms of training accuracy is provided by \vanilaFL{} thanks to benefit of unconstrained communication.
With \propImperfect{}, 93.5\,\% of average accuracy and 10.7 variance is observed.
Client scheduling utilizing all $\numberofresourceblocks=6$ RBs offers the advantage for the aforementioned performance over all the other methods considering the communication limitations.
It can be also be seen that \propPerfect{} and \basePaper{} have almost equal means (92\,\%) and variances of 16.9 and 19.5, respectively.
Scheduling clients to train over a larger dataset in \propPerfect{} provides a lower variance in accuracy over \basePaper{}.
Although \baseRand{} is CSI-agnostic, it yields an average accuracy of 92.8\,\% and the highest variance of 19.41.
This indicates that the client scheduling without any insight of datasize distribution and CSI cannot provide high training accuracy or fairness under communication constraints.

\section{Conclusion}
\label{conclu}

In this paper, we proposed a novel joint client scheduling and RB allocation policy for FL over wireless links under imperfect CSI.
The problem of client scheduling and RB allocation was cast to minimize the training loss and the CSI uncertainties. 
Resorting to GPR-based channel prediction method and deriving an upper bound for the loss of accuracy in FL compared to a centralized approach, the stochastic optimization problem was solved using Lyapunov optimization.
By means of an extensive set of simulations, we evaluated the performance of the proposed methods for both perfect and imperfect CSI.
Results show that the performance of the proposed methods outperforms state of the art client scheduling and RB allocation methods, especially when training data is unevenly distributed among clients.
Analyzing the tradeoffs between computation and communication and the impact of communication errors are potential future extensions.\balance 

\bibliography{bibref}

\begin{thebibliography}{10}

\bibitem{park2018wireless}
J.~Park, S.~Samarakoon, M.~Bennis, and M.~Debbah, ``Wireless network
  intelligence at the edge,'' {\em Proceedings of the IEEE}, vol.~107,
  pp.~2204--2239, Nov. 2019.

\bibitem{konevcny2016federated}
J.~Kone{\v{c}}n{\`y}, H.~B. McMahan, D.~Ramage, and P.~Richt{\'a}rik,
  ``Federated optimization: Distributed machine learning for on-device
  intelligence,'' {\em arXiv preprint arXiv:1610.02527}, 2016.

\bibitem{samarakoon2018federated}
S.~Samarakoon, M.~Bennis, W.~Saad, and M.~Debbah, ``Federated learning for
  ultra-reliable low-latency {V2V} communications,'' in {\em 2018 IEEE Global
  Communications Conference (GLOBECOM)}, (Abu Dabhi, UAE), pp.~1--7, IEEE, Dec.
  2018.

\bibitem{chen2018lag}
T.~Chen, G.~Giannakis, T.~Sun, and W.~Yin, ``{LAG}: Lazily aggregated gradient
  for communication-efficient distributed learning,'' in {\em Advances in
  Neural Information Processing Systems}, pp.~5050--5060, 2018.

\bibitem{nishio2019client}
T.~Nishio and R.~Yonetani, ``Client selection for federated learning with
  heterogeneous resources in mobile edge,'' in {\em 2019 IEEE International
  Conference on Communications (ICC)}, (Shanghai, China), pp.~1--7, IEEE, May
  2019.

\bibitem{yang2019scheduling}
H.~H. Yang, Z.~Liu, T.~Q. Quek, and H.~V. Poor, ``Scheduling policies for
  federated learning in wireless networks,'' {\em arXiv preprint
  arXiv:1908.06287}, 2019.

\bibitem{chen2019joint}
M.~Chen, Z.~Yang, W.~Saad, C.~Yin, H.~V. Poor, and S.~Cui, ``A joint learning
  and communications framework for federated learning over wireless networks,''
  {\em arXiv preprint arXiv:1909.07972}, 2019.

\bibitem{neely2010stochastic}
M.~J. Neely, ``Stochastic network optimization with application to
  communication and queueing systems,'' {\em Synthesis Lectures on
  Communication Networks}, vol.~3, no.~1, pp.~1--211, 2010.

\bibitem{karaca2012smart}
M.~Karaca, T.~Alpcan, and O.~Ercetin, ``Smart scheduling and feedback
  allocation over non-stationary wireless channels,'' in {\em 2012 IEEE
  International Conference on Communications (ICC)}, (Ottawa, Canada),
  pp.~6586--6590, IEEE, June 2012.

\bibitem{hiriart2004fundamentals}
J.~Hiriart-Urruty and C.~Lemar{\'e}chal, {\em Fundamentals of Convex Analysis}.
\newblock Grundlehren Text Editions, Springer Berlin Heidelberg, 2004.

\bibitem{xing2015gpr}
E.~P. Xing, {\em Advanced Gaussian Processes}.
\newblock 10-708: Probabilistic Graphical Models, University in Pittsburgh,
  Pennsylvania: Carnegie Mellon School of Computer Science, Feb. 2015.

\bibitem{PrezCruz2013GaussianPF}
F.~P{\'e}rez-Cruz, S.~V. Vaerenbergh, J.~J. Murillo-Fuentes,
  M.~L{\'a}zaro-Gredilla, and I.~Santamar{\'i}a, ``Gaussian processes for
  nonlinear signal processing: An overview of recent advances,'' {\em IEEE
  Signal Processing Magazine}, vol.~30, pp.~40--50, July 2013.

\end{thebibliography}
\bibliographystyle{ieeetr}
\end{document}